\documentclass[aps,pra,twocolumn,groupedaddress,superscriptaddress,showpacs,nofootinbib]{revtex4}
\usepackage{bbm,amsmath,amssymb,amsbsy,graphicx,subfigure}
\usepackage{hyperref,color}
\usepackage{pifont}
\usepackage[latin1]{inputenc}
\newtheorem{theorem}{\bf Theorem}

\newtheorem{definition}[theorem]{{\bf Definition}{}}
\newtheorem{lemma}[theorem]{\bf Lemma}

\newcommand{\be}{\begin{equation}}
\newcommand{\ee}{\end{equation}}
\newcommand{\ben}{\begin{eqnarray}}
\newcommand{\een}{\end{eqnarray}}
\newcommand{\bes}{\begin{subequations}}
\newcommand{\ees}{\end{subequations}}
\newcommand{\dg}{\dagger}
\newcommand{\id}{\mathbbm{1}}
\newcommand{\C}{\mathbb{C}}
\newcommand{\sse}{{\mbox{\ding{75}}}}

\newcommand{\tr}{{\rm Tr}}

\newcommand{\ket}[1]{|#1\rangle}
\newcommand{\bra}[1]{\langle#1|}

\newcommand{\eq}[1]{Eq.~(\ref{#1})}

\renewcommand{\Re}{\textrm{Re}}

\newcommand{\Salerno}{\affiliation{Dipartimento di Matematica e Informatica, Universit\`a degli Studi di
Salerno, CNISM, Unit\`a di Salerno, and INFN, Sezione di Napoli -
Gruppo Collegato di Salerno, Via Ponte don Melillo, I-84084 Fisciano (SA),
Italy}
}

\newcommand{\Nottingham}{\affiliation{School of Mathematical Sciences, University of Nottingham,
University Park,  Nottingham NG7 2RD, UK.}
}

\newcommand{\Corresponding}{\thanks{Corresponding author. Electronic address:
illuminati@sa.infn.it}
}

\begin{document}

\title{Entanglement quantification by local unitaries}

\author{A. Monras}\Salerno
\author{G. Adesso}\Nottingham
\author{S. M. Giampaolo}\Salerno
\author{G. Gualdi}\Salerno
\author{G. B. Davies}\Nottingham
\author{F. Illuminati}
\Corresponding
\Salerno

\begin{abstract}
Invariance under local unitary operations is a fundamental property that must be obeyed by every proper measure of quantum entanglement. However, this is not the only aspect of entanglement theory where local unitaries play a relevant role. In the present work we show that the application of suitable local unitary operations defines a family of bipartite entanglement monotones, collectively referred to as "mirror entanglement". They are constructed by first considering the (squared) Hilbert-Schmidt distance of the state from the set of states obtained by applying to it a given local unitary. To the action of each different local unitary there corresponds a different distance. We then minimize these distances over the sets of local unitaries with different spectra, obtaining an entire family of different entanglement monotones. We show that these mirror entanglement monotones are organized in a hierarchical structure, and we establish the conditions that need to be imposed on the spectrum of a local unitary for the associated mirror entanglement to be faithful, i.e.~to vanish on and only on separable pure states. We analyze in detail the properties of one particularly relevant member of the family, the "stellar mirror entanglement" associated to traceless local unitaries with nondegenerate spectrum and equispaced eigenvalues in the complex plane. This particular measure generalizes the original analysis of [Giampaolo and Illuminati, Phys. Rev. A {\bf 76}, 042301 (2007)], valid for qubits and qutrits. We prove that the stellar entanglement is a faithful bipartite entanglement monotone in any dimension, and that it is bounded from below by a function proportional to the linear entropy and from above by the linear entropy itself, coinciding with it in two- and three-dimensional spaces.
\end{abstract}


\date{May 16, 2011}

\maketitle

\section{Introduction}\label{secIntro}

Achieving a satisfactory understanding of the nature and structure of quantum correlations is of paramount importance in quantum information theory \cite{hororev} as well as in the study of complex quantum systems \cite{faziorev}; very recent speculations even hint at possible fundamental roles played by quantum entanglement in biological systems and processes with enhanced properties of quantum coherence \cite{biology}. For bipartite quantum systems prepared in globally pure states, there is universal consensus on the fact that quantum correlations identify with entanglement, and they can be signaled by either one of several features that distinguish them from the classical ones. For instance, their non-local character, or their operational significance in quantum informational primitives such as entanglement creation and distillation, or the performance they enable as resources for quantum communication protocols such as teleportation \cite{hororev}. Entanglement in this case is simply measured by how much information one is missing on the state of the global system by accessing only one part of it, hence capturing the correlations between the two parties. Mathematically, such an information is quantified by the {\it entropy of entanglement}, i.e.~the von Neumann entropy (or any monotonically increasing function of it) of the reduced density matrix of either of the two subsystems \cite{Bennett96}, although it is worth remarking that other inequivalent measures of pure-state bipartite entanglement can be introduced beyond the von Neumann entropy, such as the infinite set of R\'enyi entanglement entropies of the reduced density matrix (entanglement spectrum) \cite{Hastings} and the bipartite geometric entanglement \cite{Blasone}, that are particularly relevant in the investigation of ground-state properties of condensed matter systems and in the theory of quantum information with continuous variables \cite{Adesso}.

For bipartite mixed states and for general multipartite states (pure or mixed), several inequivalent entanglement monotones (as well as a number of measures of more general types of quantum correlations beyond entanglement in mixed states \cite{discord}) have been proposed, each apt to capture a different signature of quantumness and/or possessing a different operational meaning, especially in relation to different informational tasks \cite{virmani}. Whilst all entanglement monotones must vanish on separable states, there can be some that vanish also on some entangled states, if the latter fail to encode the particular resource character associated to a given entanglement measure. This is the case, for example, of bound entangled states, which have a nonzero entanglement cost but a vanishing distillable entanglement, as no entangled singlets can be extracted from them by local operations and classical communication (LOCC) in the asymptotic regime \cite{boundent}. Fundamental requirements for any {\it bona fide} entanglement monotone are then monotonicity under LOCC -- operations that cannot increase bipartite entanglement on average -- and invariance under local unitary (LU) operations \cite{vidal}. The latter is in fact a basic requirement for any measure of correlation in general, since the choice of the local basis in which the density matrix
of a system is expressed cannot of course affect the information encoded in shared correlations between two (or more) subsystems. This fact has motivated the search for the simplest LU-invariant ``normal forms'' of quantum states in discrete as well as continuous variable systems \cite{lindenetal}, so as to reduce as much as possible the number of state parameters needed for a complete evaluation of a particular entanglement monotone.

Invariance of entanglement under LUs has however more far-reaching consequences, especially, and not entirely without surprise, in its quantification. Given a generic state (pure or mixed) of a bipartite systems, Li-Bin Fu investigated the consequences of applying a local cyclic operation on one of the subsystems \cite{Fu}. Fu denoted by local cyclic operation any LU that leaves the corresponding reduced state invariant. Consider a bipartite quantum system $(A|B)$ in a global pure state $\ket{\Psi}$ with reduced density matrices $\rho_A$ and $\rho_B$ respectively for subsystems $A$ and $B$, and denote by $U_A$ a LU acting on $\rho_A$ only.
Then, the LU $U_A$ is cyclic if $[U_A,\rho_A] = 0$.
Although being local, thus leaving the entanglement unchanged, such an operation changes the global state, yielding a nonlocal effect that can be detected only by measuring the two subsystems jointly. Employing the fidelity induced by the Hilbert-Schmidt metric, Fu identified such a nonlocal
effect by the distance between the initial and the final state. Recently, Fu's pioneering work has been greatly extended by Gharibian, Kampermann, and Bruss \cite{Bruss}. They focused on the {\emph maximization} of the Fu distance for bipartite states in Hilbert spaces of arbitrary dimension as a possible indicator of non-local properties, and derived and discussed closed formulae for the maximal Fu distance in three relevant cases: (pseudo)pure quantum states, Werner states, and two-qubit states. In between, two of us (S.M.G. and F.I.) investigated independently the consequences of LUs on global pure states of $2 \times D$ and $3 \times D$ bipartite systems \cite{GiampaoloIlluminati}. They attacked the problem from the opposite side, investigating the {\emph minimization} of the (squared) Fu distance, and proved, somewhat surprisingly, that in these two particular cases the minimum (squared) Fu distance is a full bipartite entanglement monotone, coinciding with the linear entropy of entanglement (also known as tangle for qubit systems \cite{ckw,osborne}). A similar analysis was performed to define a geometric LU-based entanglement measure for Gaussian states of continuous variable systems, where subsystem $A$ comprises a single bosonic mode \cite{squoCV}.

In the present paper we will present a complete generalization of the analysis carried out in Ref. \cite{GiampaoloIlluminati} to all pure states of bipartite quantum systems with Hilbert space
${\cal H_{AB}}$ of arbitrary finite dimension. We will show that one can construct an entire family of bipartite entanglement monotones that capture quantum correlations as quantified by the action of minimally perturbing LUs on global pure states. Specifically, we prove that the (squared) Hilbert-Schmidt distance between a pure bipartite state $\ket{\psi}_{AB} \in {\cal H_{AB}} = {\cal H}_A \otimes {\cal H}_B$ and the pure state $U_A\ket{\psi}_{AB}$ obtained from it by applying a LU operation $U_A$ on subsystem $A$ only, once suitably optimized (minimized) over all LUs with fixed spectrum, defines a hierarchy of bipartite entanglement monotones. Moreover, the cyclic condition $[U_A,\rho_A] = 0$, rather than being imposed {\emph a priori} is derived as a consequence of the minimization (optimization) procedure. We denote any such pure-state bipartite entanglement monotone as ``{\em mirror entanglement}''. This is pictorially reminiscent of someone mirroring herself/himself in a mirror (which is curved, symbolizing the action of a LU): in absence of entanglement, the mirror image of the original pure state under a LU (the mirror) is a perfect reflection. Viceversa, the more entanglement is contained in the state of the system, the more distorted is the image that is reflected by the mirror.

As just stated, no constraints on the employed LUs need to be imposed {\emph ab initio}: all the mirror entanglement measures are proven to be LOCC-monotones for arbitrary pure states of bipartite systems in any dimension. However, we will show that by imposing, as originally done in Ref. \cite{GiampaoloIlluminati}, specific requirements on $U_A$, namely a fully nondegenerate spectrum, one restricts the class further to {\it faithful} mirror entanglement monotones that vanish if and only if the state $\ket{\psi}_{AB}$ is a product state. Upon further restricting the admissible LUs by requiring that they be traceless and with equispaced, nondegenerate eigenvalues (thus with a pattern resembling a star in the complex plane), we
will single out a special faithful mirror entanglement monotone, that we will name ``{\em stellar mirror entanglement}'' . We will prove that the stellar mirror entanglement enjoys the property of being a lower bound to the linear entropy in any dimension, reducing to the latter in the special cases ${\cal H}_A=\C^2$ and ${\cal H}_A=\C^3$ originally considered in \cite{GiampaoloIlluminati}. Moreover, it is an upper bound to a function proportional to the linear entropy, with the proportionality constant being a simple function of the dimension of the reduced Hilbert space ${\cal H}_A$.

We remark that, by construction, the class of mirror entanglement measures is experimentally accessible by means of interferometric schemes \cite{interfero} involving at least two copies of a given bipartite state, one of which needs to be rotated by suitable LUs. In principle, each mirror entanglement monotone can be straightforwardly extended to mixed bipartite states via the conventional convex roof construction. Solving the convex roof optimization problem is of course in general a formidable task. However, as the mirror and stellar entanglement measures are defined in terms of distances (in partial analogy with the construction of the geometric measures of entanglement \cite{Wei,Blasone}), in the conclusions we will briefly discuss how it might be possible to envisage alternative strategies to compute their mixed-state extension without resorting directly to the convex roof construction.

The paper is organized as follows. In Sec.~\ref{secAlex} we define the class of mirror entanglement measures for all pure states of bipartite quantum systems, prove their monotonicity under LOCC, characterize their hierarchic structure, and determine the conditions for faithfulness. In Sec.~\ref{secGary} we focus on the stellar mirror entanglement and investigate its relationship with the linear entropy of entanglement, providing the exact lower and upper bounds that relate the two quantities (The detailed proofs of the bounds are reported in two Appendices). Finally, in Sec.~\ref{secDiscuss} we briefly discuss some of the implications of our results and some possible future research directions concerning the extension to mixed states and the problem of identifying total and partial pure-state factorization (separability) in quantum many-body systems.

\section{mirror entanglement: definition and monotonicity}\label{secAlex}

Let us consider a bipartite system in a pure quantum state $\ket{\psi}\equiv\ket{\psi}_{AB}$ belonging to a Hilbert space ${\cal{H}}_{AB} = {\cal H}_A \otimes {\cal H}_B \equiv \C^{d_A} \otimes \C^{d_B}$. We will assume without loss of generality that $d \equiv d_A \le d_B$.
Let us consider general  LUs acting on ${\cal H}_A$  of the form
\begin{equation}\label{squdos}
	W_{\Lambda,A} \equiv W_{\Lambda} = \sum_j \lambda_j \ket{\phi_j}\bra{\phi_j},
\end{equation}
where
\begin{equation}\label{lambda}
\Lambda = \{\lambda_j\}\equiv\{e^{i\theta_j}\} \quad (j=1,\ldots,d)\,,
\end{equation} denotes the spectrum of the eigenvalues of $W_{\Lambda}$.
The maximal fidelity (squared overlap) between state $\ket\psi$ and the LU-transformed state $W_{\Lambda}\ket\psi$ is
\begin{equation}\label{Fpsi}
	F^\Lambda_\psi=\max_{W_{\Lambda}}|\bra{\psi}W_{\Lambda}\ket{\psi}|^2 \; ,
\end{equation}
and it takes non-negative real values in the interval $[0,1]$.

\begin{definition}[mirror Entanglement]\label{def1}
The bipartite $\Lambda$-{\em mirror entanglement} ($\Lambda$ME) between subsystems $A$ and $B$ in the state $\ket\psi$ is defined as the square of the minimum Euclidean distance between $\ket\psi$ and the set of transformed states $W_{\Lambda}\ket\psi$ obtained by the action of LUs of the form \eq{squdos} with spectrum $\Lambda$ on subsystem $A$:
\begin{equation}\label{se}
{\cal E}_\Lambda(\psi) \doteq \min_{W_{\Lambda}} \left(1-|\bra{\psi}W_{\Lambda}\ket{\psi}|^2\right) = 1-F^\Lambda_\psi \; .
\end{equation}
\end{definition}

Consider the reduced density matrix of subsystem $A$:
\begin{equation}
\varrho \equiv \varrho_A=\tr_B[\ket\psi\bra\psi] \; .
\label{Reduced}
\end{equation}
Definition (\ref{se}) is recast in terms of the reduced state (\ref{Reduced})
by rewriting Eq. (\ref{Fpsi}) as
\begin{align*}
	F^\Lambda_\psi&=\max_{W_{\Lambda}} |\tr[{W_{\Lambda}} \ket\psi\bra\psi]|^2\\&=\max_{W_{\Lambda}} |\tr_A[{W_{\Lambda}}\,\tr_B[\ket\psi\bra\psi]]|^2\\&=\max_{W_{\Lambda}}|\tr[{W_{\Lambda}}\varrho]|^2
\; .
\end{align*}
By the monotonicity of the square function one can write $\sqrt {F^\Lambda_\psi}=\max_{W_{\Lambda}} |\tr[W_{\Lambda}\varrho]|$.
Let $\ket i$ be the eigenbasis of $\varrho$ and $p_i$ its eigenvalues, so that one has the spectral decomposition $\varrho=\sum_i p_i \ket i\bra i$. The set of allowed LUs, that is, the set of unitary matrices acting on ${\cal H}_A$ with spectrum $\Lambda$ [\eq{lambda}], can be written in terms of $V_\Lambda=\sum_i \lambda_i\ket i\bra i$ as
 \begin{equation}\label{WUV}
 	{W_{\Lambda}}=U V_\Lambda U^\dagger=\sum_i \lambda_i U\ket i\bra iU^\dagger \; ,
\end{equation}
where $U$ rotates the eigenbasis of $\varrho$ into the eigenbasis of ${W_{\Lambda}}$, $\ket{\phi_i}=U\ket i$.
In principle, $U$ can be any $SU(d)$ unitary matrix. We can write $\sqrt{F^\Lambda_\psi}$ as
\begin{eqnarray}\label{expr1}
	\sqrt{F^\Lambda_\psi}&=&\max_{U\in SU(d)} \left|\tr\left[U{V_\Lambda}U^\dagger \varrho\right]\right| \nonumber\\
	&=&\max_{U\in SU(d)}\bigg|\sum_i \lambda_i\,\tr\big[U\ket i\bra iU^\dagger \sum_j p_j \ket j\bra j\big]  \bigg| \nonumber\\
	&=&\max_{U\in SU(d)}\bigg|\sum_i \lambda_i\sum_j p_j |u_{ij}|^2 \bigg| \; ,
\end{eqnarray}
where $u_{ij}=\bra{i}U\ket{j}$.

We will now show (Theorems~\ref{TVanSep} and~\ref{TLOCC} below) that the $\Lambda$ME measures
are indeed legitimate pure-state entanglement monotones.

\begin{theorem}\label{TVanSep} The ME vanishes on pure separable (i.e., product) bipartite states $\ket{\psi^\otimes} = \ket{\psi_A}\otimes \ket{\psi_B}$.\end{theorem}
{\it Proof.}
For a product state, $\varrho$ is a rank-one matrix, with eigenvalues $p_j= \delta_{jk}$ for some index $k$. Choosing $U=\id$ one has
$|\sum_i \lambda_i\sum_j p_j |u_{ij}|^2| = |\sum_i \lambda_i\sum_j \delta_{jk} \delta_{ij}|=|\lambda_k|=|e^{i \theta_k}|=1\equiv\sqrt{F^\Lambda_{\psi^{\otimes}}}$. Therefore, from \eq{se} ${\cal E}_\Lambda(\psi^\otimes)=0$ for any $\Lambda$. \hfill $\Box$

Before tackling the monotonicity of the ME under LOCC, we first prove an auxiliary lemma that simplifies the optimization problem involved in the definition of the ME.
\begin{lemma}\label{lemmaPerm} The maximizing unitary $U$ in \eq{expr1} is a permutation matrix.\end{lemma}
{\it Proof.} Eq.~\eqref{expr1} can be written as
\[	\sqrt{F^\Lambda_\psi}=\max_{U\in SU(d)}\left|\tr[M^\Lambda_\psi B(U)]\right|\,,
\]
where $[B(U)]_{ij}=|u_{ij}|^2$ and $(M^\Lambda_\psi)_{ij}=p_i \lambda_j$. Noticing that $B(U)$ is a unistochastic matrix, we can write
\[
	\sqrt{F_\psi^\Lambda}=\max_{B\ \!\textrm{unistoch.}}\left|\tr[M^\Lambda_\psi B]\right|\leq\max_{B\in \textsf{B}_d}\left|\tr[M^\Lambda_\psi B]\right|\,,
\]
where we have enlarged the optimization domain to the whole set $\textsf{B}_d$  of all  $d\times d$ doubly stochastic matrices, i.e.~the $d$-dimensional Birkhoff polytope. By the Birkhoff-von Neumann theorem,  $\textsf{B}_d$ is the convex hull of the set $\textsf{S}_d$ of $d \times d$ permutation matrices (that is, the permutation matrices in $\textsf{S}_d$ are the extreme points of $\textsf{B}_d$). We can thus write  $B=\sum_k q_k S_k$, where $S_k \in \textsf{S}_d$, and $\vec q = \{q_k\}$ is a $d!$-dimensional probability vector. 
The maximal fidelity becomes
$\sqrt{F^\Lambda_\psi}\leq\max_{\vec q}\left|\sum_k q_k\tr[M^\Lambda_\psi S_k]\right|
\leq \max_{\vec q}\sum_k q_k \left|\tr[M^\Lambda_\psi S_k]\right|$, where we have used the triangle inequality.
Let $S_{\max}$ be the permutation matrix that maximizes $\left|\tr[M^\Lambda_\psi S]\right|$. Then
$\sqrt{F^\Lambda_\psi} \leq  \left(\max_{\vec q}\sum_k q_k\right) \left|\tr[M^\Lambda_\psi S_{\max}]\right|=\left|\tr[M^\Lambda_\psi S_{\max}]\right|$.
We are left to show that $S_{\max}=B(U)$ for some $U$. This is achieved by noticing that all permutation matrices, including $S_{\max}$, are orthogonal and hence unitary, and that $B(S_{\max})=S_{\max}$. This concludes the proof. \hfill{$\Box$}

As a corollary of Lemma \ref{lemmaPerm}, we find that the optimal LU operation ${W_{\Lambda}}=S_{\max}V_{\Lambda}S_{\max}^\dagger$ that maximizes $F^\Lambda_\psi$ [\eq{Fpsi}]
commutes with the reduced state $\varrho$. To see this, let us write
$S_{\max}=\sum_i\ket{{\sigma_i}}\bra{i}$,
 where $\sigma$ is the permutation described by the matrix $S_{\max}$. Then
\begin{eqnarray}
\label{squo}
	{W_{\Lambda}}&=&\left(\sum_{i}\ket{{\sigma_i}}\bra{i}\right)\,V_\Lambda\,\left(\sum_j \ket{j}\bra{\sigma_j}\right) \nonumber\\
	&=&\sum_{i}\lambda_i\,\ket{{\sigma_i}}\bra{{\sigma_i}}\\
	&=&\sum_{i}\lambda_{\sigma_i^{-1}}\,\ket{i}\bra{i} \nonumber \; ,
\end{eqnarray}
which is diagonal in the same basis as $\varrho$, and therefore $[\varrho,{W_{\Lambda}}]=0$. The last result shows that the eigenvectors of the optimal LU $W_{\Lambda}$ that solves the minimization in the definition of the $\Lambda$ME [\eq{se}] are just obtained by a reordering of the eigenvectors of the reduced state $\varrho$. Collecting the previous findings, $\sqrt{F^\Lambda_\psi}$ can be written as
\begin{equation}\label{eq:permutation_form}
	\sqrt{F^\Lambda_\psi}=\max_{S\in \textsf{S}_d}\left|\tr[M^\Lambda_\psi S]\right|\,.
\end{equation}

We are now ready to prove the following important result.

\begin{theorem}\label{TLOCC} The ME is monotonically non-increasing under LOCC operations, i.e., is a full pure-state bipartite entanglement monotone.\end{theorem}
{\it Proof.} We will prove\footnote{Here and in the following steps of the proof we omit the label $\Lambda$ for ease of notation.} that $F_\psi$ is monotonically increasing under LOCC, which implies the statement.
An arbitrary pure state $\ket{\psi}$ is ensemble-transformed under LOCC according to: $\ket{\psi}\rightarrow \{p_i,\ket{\psi_i}\}$, where each LOCC-transformed state of the ensemble reads:
\begin{equation}
	\sqrt{p_i} \ket{\psi_i}=(A_i\otimes\openone_B) \ket{\psi} \; .
\end{equation}
The positive weights $\{ p_i \}$ satisfy the normalization condition $\sum_{i} p_i = 1$,
while the Kraus operators associated to the local dynamics satisfy the POVM (positive operator
valued measure) completeness relation: $\sum_i A_i^\dagger A_i=\openone$. Let the reduced state be $\rho=\tr_B[\ket\psi\bra\psi]$ and likewise the reduced LOCC-transformed states be $\rho_i=\tr_B[\ket{\psi_i}\bra{\psi_i}]$. For each of the latter, the local dynamics yields:
\begin{equation}
	p_i\rho_i=A_i\rho A_i^\dagger \; .
\end{equation}
Then, in order to prove that ${\cal E}(\psi) \geq \sum_{i} p_i {\cal E}(\psi_i)$, it is sufficient
to prove that $\sum_i p_i\sqrt{F(\psi_i)}\geq\sqrt{F(\psi)}$. Let
\begin{equation}\label{polar}
	A_i \sqrt\rho=\sqrt{A_i\rho A_i^\dagger}V_i=\sqrt{p_i}\sqrt\rho_i V_i
\end{equation}
be the polar decomposition of $A_i \sqrt\rho$, where $V_i$ is a suitable unitary matrix.
Exploiting the properties of $F$, we can write:
\begin{align}
\nonumber
	\sum_i p_i\sqrt{F(\psi_i)}=&\sum_ip_i \max_{W_i}\left|\tr[\rho_i W_i]\right|\\
\nonumber
	=&\sum_ip_i \max_{W_i}\left|\tr[\rho_i V_i W_iV_i^\dagger]\right|\\
\nonumber
	=&\sum_i p_i\max_{W_i}\left|\tr[(V_i^\dagger\sqrt\rho_i)(\sqrt\rho_i V_i) W_i]\right|\\
\nonumber
	=&\sum_i\max_{W_i}\left|\tr[(\sqrt\rho A_i^\dagger)( A_i \sqrt\rho) W_i]\right|\\
\nonumber
	\geq&\max_{W}\sum_i\left|\tr[\sqrt\rho A_i^\dagger A_i \sqrt\rho W]\right|\\
\nonumber
	\geq&\max_{W}\left|\tr[\sqrt\rho\left(\sum_i A_i^\dagger A_i \right)\sqrt\rho W]\right|\\
\nonumber
	\geq&\max_{W}\left|\tr[\rho W]\right|\\
\label{line8}
	=&\sqrt{F(\psi)}.
\end{align}
%
This concludes the proof that local operations on party $A$ do not increase $1-F$.
The proof for operations on party $B$ follows trivially. We have:
\begin{align}
\nonumber
	\sum_i p_i\sqrt{F(\psi_i)}=&\sum_ip_i \max_{W_i}\left|\bra{\psi_i} W_i\otimes\openone\ket{\psi_i}\right|\\
\nonumber
	=&\sum_i\max_{W_i}\left|\bra\psi W_i\otimes B^\dagger_iB_i\ket\psi\right|\\
\nonumber
\geq &\max_W\left|\bra{\psi} W\otimes\sum_iB^\dagger_iB_i\ket{\psi}\right|\\
\label{line14}
=&\sqrt{F(\psi)} \; .
\end{align}
Therefore, since ${\cal E}(\psi) = 1-F_\psi$ and all local operations do not increase $1-F$,
we have proved that the ME is non-increasing under LOCC.
\hfill $\Box$

We have thus introduced a family of pure-state bipartite entanglement monotones, that satisfy the three fundamental axiomatic properties of vanishing on separable states, being invariant under local unitaries, and being monotonic under LOCC \cite{Nielsen,vidal,hororev,virmani,geometry}. They are associated to the minimum distance between a quantum state $\ket\psi$ and its image after suitable unitary operations with fixed spectrum performed on one subsystem only. Surprisingly enough, one sees that a LU operation on one part of a bipartite system, while leaving the entanglement invariant, leads nevertheless to a state alteration whose proper quantification is itself an entanglement measure. The properties of the spectrum $\Lambda$ define the shape, or reflectivity, of a fictitious mirror which produces the image of the quantum state after the action of a LU. In the absence of entanglement, there exists always one such LU that leaves the state invariant, yielding a perfect reflection from the mirror. If the state $\ket{\psi}$ is entangled, the action of the minimal or least perturbing LU with spectrum $\Lambda$ necessarily results in a transformed state which has a nonmaximal fidelity with $\ket{\psi}$; in turn, this distortion quantifies the amount of bipartite entanglement present in $\ket{\psi}$.

The class of $\Lambda$ME exhibits a hierarchical structure depending on the characteristics of the spectrum $\Lambda=\{e^{i \theta_j}\}$. One extreme case is represented by $\theta_j=0$ $\forall j$. In this case the identity is clearly the extremal LU operation that defines the $\id$ME according to \eq{se}, and the ensuing entanglement measure ${\cal E}_\id(\psi)$ is trivially zero for all quantum states $\ket{\psi}$. Progressing in the hierarchy, if $\Lambda$ contains an $r$-degenerate eigenvalue, then the corresponding ME vanishes on entangled states $\ket{\psi}$ of Schmidt rank $\leq r$. On the opposite extreme, if the eigenvalues in $\Lambda$ are all nondegenerate, one obtains the most sensitive measure of mirror entanglement $\Lambda$ME. This classification is summarized by the following theorem:

\begin{theorem}\label{teorango}
Let $W_{\Lambda}$ be a LU and let the associated spectrum $\Lambda$ have degeneracy $r$ (any eigenvalue repeated at most $r$ times). Then ${\cal E}_\Lambda(\psi)=0$ if and only if the Schmidt rank of $\ket\psi$, $\mathrm{SR}(\psi)$, is no larger than $r$.
\end{theorem}
{\it Proof.} The Schmidt rank of $\ket\psi$ amounts to the rank of the reduced density matrix $\varrho$, or the number of nonvanishing elements in the probability vector $\vec p$. We first prove sufficiency and then necessity:
\begin{description}
\item I) [$\mathrm{SR}(\psi)\leq r\Rightarrow {\cal E}_\Lambda(\psi)=0 $]. \quad \\ If the Schmidt rank of $\psi$ is $s\leq r$ then $|\tr[S V_\Lambda S^\dagger \varrho]|=1$ can be attained by any permutation matrix $S_{\max}$ such that $S_{\max}^\dg$ maps the $s$-dimensional domain of $\varrho$ into the $r$-fold degenerate subspace of $V_\Lambda$, so that $|\tr[S V_\Lambda S^\dagger \varrho]|=|\sum_{i=1}^s \lambda_i p_i|=1$, where $\lambda_1=\ldots=\lambda_r$. In this way one has ${\cal E}_\Lambda(\psi)=0$.

\item II) [${\cal E}_\Lambda(\psi)=0 \Rightarrow \mathrm{SR}(\psi)\leq r $]. \quad \\  To prove the inverse implication, let $\sigma$ be the maximizing permutation in Eq.~\eqref{eq:permutation_form}, $\bar\lambda_i=\lambda_{\sigma^{-1}_i}$, and let $\Sigma$ be the set of indices for which $p_i\neq0$. Then $s=|\Sigma|$. Expanding $F_\psi^\Lambda$ we get
\begin{subequations}
\begin{eqnarray*}
	F_\psi^\Lambda&=&\Big| \sum_{i\in \Sigma} \bar\lambda_i\, p_i\Big|^2\\
	&=&\sum_{i\in \Sigma} p_i^2+\sum_{\{i\neq j\}\in \Sigma} \bar\lambda_i^* \bar\lambda_jp_i p_j\\
	&=&\sum_{i\in \Sigma} p_i^2+\sum_{\{i<j\}\in \Sigma}(\bar\lambda_i^* \bar\lambda_j+ \bar\lambda_j^* \bar\lambda_i)p_i p_j\\
	&\leq&\sum_{i\in \Sigma} p_i^2+2\sum_{\{i<j\}\in \Sigma}p_i p_j\\
	&=&\Big(\sum_{i\in \Sigma} p_i\Big)^2=1 \; .
\end{eqnarray*}
\end{subequations}
The inequality follows from $(a^* b+b^* a)\leq 2|a||b|$, and it is saturated if and only if $\Re[a^*b]=1$, thus for each pair $i,j\in\Sigma$ we have
\begin{equation}
	\bar\lambda_i=\bar\lambda_j \quad\forall i,j\in\Sigma \; .
\end{equation}
By assumption, there are at most $r$ equal values of $\bar\lambda$, thus $|\Sigma|\leq r$. This completes the proof.\hfill$\Box$
\end{description}

\begin{figure}[tb]
	\includegraphics[width=.35\textwidth]{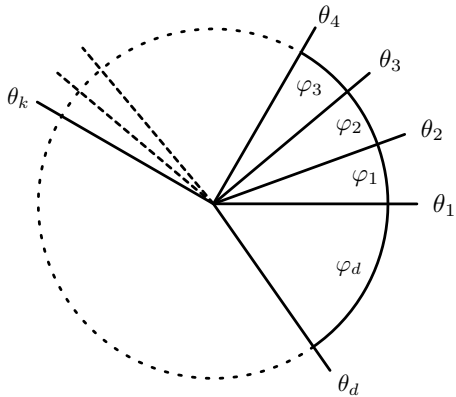}\caption{\label{fig:phases} Parameterization of the spectrum $\Lambda$ of the LU $W_{\Lambda}$, given by the normalized phases $\theta_i$ [\eq{lambda}] or the gaps $\varphi_i$ between phases (in units of $2\pi$rads). A spectrum is degenerate whenever a $\varphi_k$ equals zero, which corresponds to the boundary of the simplex [see Fig.~\ref{fig:simplex}].}
\end{figure}

\begin{figure}[t]
	\includegraphics[width=.4\textwidth]{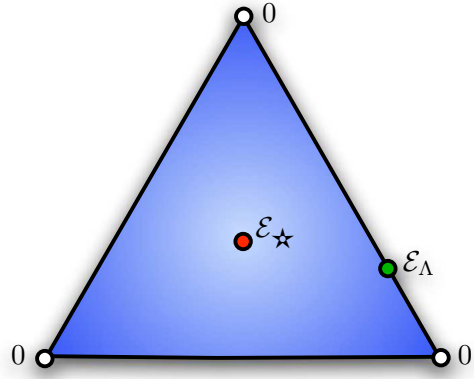}\caption{\label{fig:simplex} (Color online) Representation of the set of $\Lambda$-mirror entanglement monotones for $d=3$ in the three-dimensional simplex of the gaps $\{\varphi_k\}$ associated to the spectrum of $\Lambda$. The stellar monotone ${\cal E}_\sse$ is placed at the center of the simplex ($\varphi_i=1/3$) and denoted by a full (red) circle.
The trivial monotones, i.e. those that are identically vanishing, are associated to spectra for which two out of three $\varphi_i$'s are zero. These monotones fall at the extremal points (corners) of the simplex and are represented by full (white) circles (white dots). Partially degenerate monotones ${\cal E}_\Lambda$ are those for which one or more $\varphi_i$ is vanishing. They fall at the boundaries of the simplex. For instance, the particular case reported by a full (green) circle on one side of the simplex represents a partially degenerate monotone that is nonvanishing only on entangled states with Schmidt rank SR=3.}
\end{figure}

By proper parameterization one can obtain a pictorial representation of all the $\Lambda$ME monotones. To do so, consider again the LU spectrum $\Lambda=\{e^{i\theta_i}\}$. The $\Lambda$ME is clearly invariant under permutations of the phases  $\theta_i$, therefore one can restrict to $0\leq\theta_1\leq \theta_2\leq\ldots\leq\theta_d\leq2\pi$ without loss of generality. Furthermore, every monotone ${\cal E}_\Lambda$ is invariant under global phase shifts of the eigenvalues, thus $\theta_1=0$ can be assumed throughout. Finally, one can equivalently re-parameterize the phases by the set of gaps $\varphi_k=(\theta_{k+1}-\theta_k)/2\pi$, with $\varphi_d=1-\theta_d/2\pi$. Since the phases are ordered, one has $\varphi_k\geq0$ and clearly $\sum_{k=1}^d\varphi_k=1$ [see Fig.~\ref{fig:phases}]. Thus, each $\Lambda$ME is in correspondence with a $d$-dimensional probability distribution $\{\varphi_k\}$. This means that all the $\Lambda$ME monotones can be represented in a simplex, as shown in Fig.~\ref{fig:simplex}.

As a Corollary of Theorem \ref{teorango}, one can easily identify all the $\Lambda$ME monotones which are {\em faithful},  i.e.,  vanish on and only on product states. These correspond to fully nondegenerate spectra $\Lambda$, and fill the entire shaded region in Fig.~\ref{fig:simplex}.
A particular case of a faithful measure is obtained when the eigenvalues constituting the LU spectrum $\Lambda$ are equispaced in the complex plane and add up to zero, as it is the case, e.g., when they are taken to be the $d$th roots of $(-1)^{d-1}$. Since their representation in the Argand diagram resembles a star (or a regular polygon), we name the associated entanglement monotone ``stellar mirror entanglement'' ($\sse$ME), or ``stellar entanglement" for brevity. The properties of this particularly important monotone are discussed in the forthcoming Section \ref{secGary}.

For every {\it mixed} state $\varrho_{AB}$ of a bipartite quantum system, and for any $\Lambda$, the $\Lambda$ME can be defined via the convex roof construction
\begin{equation}\label{croof}
{\cal E}_\Lambda(\varrho_{AB}) \doteq \inf_{\{p_i,\psi_i\}} p_i {\cal E}_\Lambda(\psi_i)\,,
\end{equation}
where the infimum runs over all pure-state decompositions of $\varrho_{AB}= \sum_i p_i \ket{\psi_i}\!\bra{\psi_i}$.
The convex roof construction ensures, by definition, that the ME is still a full LOCC monotone, and hence a valid measure of entanglement, even in the general case of mixed states. It is obvious that, like in the case of most other measures of entanglement, the actual computation of the convex roof on generic mixed quantum states is a formidable task. In the conclusions we will briefly discuss how the ME measures, being defined in terms of geometric distances in Hilbert space, might be extended to the general mixed-state case via alternative procedures that bypass and avoid the explicit evaluation of the convex roof construction.

\section{Stellar mirror entanglement}\label{secGary}
In this Section we study in more detail a special instance of the $\Lambda$ME monotones introduced in the previous Section, characterized by a LU with spectrum $\Lambda$ [\eq{lambda}] obeying precise constraints, defined in the following. Observing that the LUs entering in the optimization for the $\Lambda$ME
can always be represented as ${W_{\Lambda}}=U V_\Lambda U^\dagger$, as in \eq{WUV}, we then define the stellar spectrum $\Lambda=\sse$ as the one such that
\begin{equation}
	V_\sse=\exp\left[i \frac{2\pi}{d} \hat{S}_z\right]\,,
\end{equation}
where the matrix $\hat{S}_z$ represents the $z$ component of the spin-$J$ operator with $J=(d-1)/2$, $\hat{S}_z = \text{diag}\{J, J-1, \ldots, -(J-1), -J\}$. Explicitly, the stellar spectrum is given by the diagonal of $V_\sse$,
\begin{equation}\label{stellarspec}
\sse=\{e^{i \theta_j}\} \; , \quad \theta_j = \frac{d-2j+1}{d}\pi \quad (j=1,\ldots,d) \, .
\end{equation}

\begin{definition}[\label{defsse}Stellar mirror Entanglement]
The {\em stellar mirror entanglement} ($\sse$ME) ${\cal E}_\sse(\psi)$ is defined as the $\Lambda$ME [see Definition~\ref{def1}, Eq. (\ref{se})] with $\Lambda=\sse$.
\end{definition}

Reminding that the local dimension $d$ is defined as $d=\min\{d_A,d_B\}$, the eigenvalues in the stellar spectrum correspond to the $d$th complex roots of $(-1)^{d-1}$, do not give rise to any degeneracy, and are equispaced in the complex plane (resembling a star once connected by rays to the origin). It is straightforward to show that $\sum_{j=1}^d e^{i \theta_j} = 0$; as a consequence, all the LUs $W_{\sse}$ with spectrum $\sse$ are traceless: $\tr[W_{\sse}]=0$. According to Theorem \ref{teorango} proved above, the stellar mirror entanglement (or, for brevity, stellar entanglement) ${\cal{E}}_{\sse}(\psi)$ vanishes if and only if $\ket{\psi}$ is a separable (product) state; thus, the $\sse$ME is a faithful entanglement monotone. This result includes and extends to any dimension $d$ the corresponding finding of Ref.~\cite{GiampaoloIlluminati}, valid for $d=2,3$. It also generalizes to any dimension the LU-based separability criterion that was established in that same paper for $d=2,3$. Indeed,
consider the optimal LU $W_{\sse}^{opt}$, i.e. the Single-QUDit Unitary Operation (SQUDUO) that realizes the stellar entanglement by minimizing the (squared) Euclidean distance over the entire set of LUs with stellar spectrum:
${\cal{E}}_{\sse}(\psi) = \min_{W_{\sse}}\left(1-|\bra{\psi}W_{\sse}\ket{\psi}|^{2}\right)$.
It is straightforward to show that the faithfulness of the stellar entanglement ${\cal{E}}_{\sse}(\psi)$ implies the following LU-based separability criterion: A pure state $\ket{\psi}$ of a bipartite system is separable (product) if and only if the optimal SQUDUO $W_{\sse}^{opt}$ leaves it invariant:
$W_{\sse}^{opt}\ket{\psi} = \ket{\psi}$.
The faithfulness of what we here call $\sse$ME, when restricted to the qubit case $d=2$  \cite{GiampaoloIlluminati}, has played a key role in the development of a general theory for the exact and rigorous detection and characterization of fully factorized ground states in several classes of non-exactly solvable spin-$\frac12$ models on translationally-invariant lattices \cite{ourfactoriz1} as well as in more general geometries \cite{ourfactoriz2} with arbitrary spatial dimensions, both at finite size and in the thermodynamic limit. Based on the general results of the present work, in the concluding Section \ref{secDiscuss} we will discuss some possible guidelines for the extensions of such methods to the problem of the occurrence of total and partial factorizations (such as dimerization, trimerization, and polymerization) in models with local spin variables of arbitrary dimension.

\begin{figure}[tb]
\includegraphics[width=7.5cm]{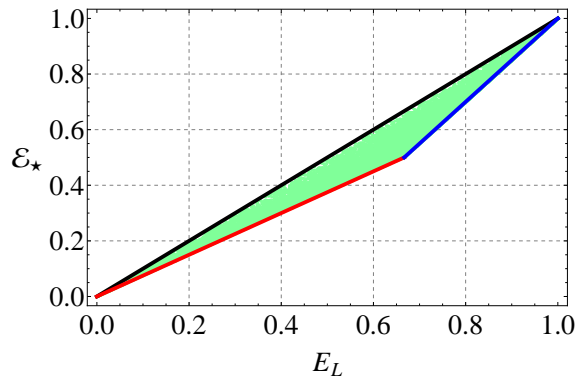}
\caption{(Color online) Behavior of the stellar entanglement ${\cal E}_\sse$ plotted against the linear entropy of entanglement $E_L$ for 20000 random pure bipartite states $\psi \in \C^{d_A} \otimes \C^{d_B}$ with local dimension $d=\min\{d_A,d_B\}=4$. The upper boundary (black online) is given by the bisectrix ${\cal E}_\sse=E_L$. These states, that accommodate for a stellar entanglement coinciding with the linear entropy are characterized by a three-fold degenerate eigenvalue in the spectrum of the reduced state $\varrho$: $p_1=p_2=p_3=\frac{p}{3}$, $p_4=1-p$ (with $0 \le p \le 1$). The lower boundary is branched into two different segments. For $0 \le E_L \le 2/3$, the minimum ${\cal E}_\sse$ satisfies ${\cal E}_\sse = \frac{3}{4} E_L$; this boundary (red online) accommodates states $\ket\psi$ with rank-2 marginal states $\varrho$: $p_1=p$, $p_2=1-p$, $p_3=p_4=0$. The second branch (blue online) of the lower boundary, defined for $2/3 < E_L \le 1$, accommodates states satisfying ${\cal E}_\sse=\frac32 E_L-\frac12$. These states have reduced density matrix $\varrho$ with doubly degenerate spectrum of the form: $p_1=p_2 = \frac{1+p}{4}$, $p_3=p_4=\frac{1-p}{4}$. All the quantities plotted are dimensionless.}
\label{figrandello}
\end{figure}

At this stage, we notice that by using 
the results of Lemma~\ref{lemmaPerm}, according to which the optimal change-of-basis matrix $U$ is a permutation matrix, we can write the following compact expression for the $\sse$ME:
\begin{equation}\label{ssecomp}
{\cal E}_\sse(\psi) = \min_\sigma \left(1-  \sum_{i,j=1}^d \cos\left[\frac{2\pi(i-j)}{d}\right]  p_{\sigma_i} p_{\sigma_j}\right) \,,
\end{equation}
where the optimization is over all permutations $\sigma$ encoding a reordering of the eigenvalues $p_k$ of the reduced state $\varrho$.
Equipped with this expression, we can proceed to investigate the relation of the $\sse$ME with other
measures of entanglement that are expressed as sums of products of eigenvalues of the reduced density matrix. The foremost measure of this kind is the {\it linear entropy of entanglement} $E_L(\psi)$, defined as the linear entropy of the reduced density matrix $\varrho$ \cite{geometry}:
\begin{equation}\label{tangle}
E_L(\psi) = S_L(\varrho) \equiv \frac{d}{d-1}(1-\tr[\varrho^2]) = \frac{d}{d-1}\bigg(1-\sum_{i=1}^d p_i^2\bigg)\!.
\end{equation}
It has been observed in Ref.~\cite{GiampaoloIlluminati} that the $\sse$ME and the linear entropy of entanglement coincide exactly on all pure states $\ket\psi$ of bipartite systems with reduced states $\varrho$ of local dimension $d=2$ or $d=3$ (qubit or qutrit). This can be easily verified by comparing Eqs.~\eqref{ssecomp} and \eqref{tangle} and recalling the normalization condition $\tr\varrho=\sum_i p_i = 1$. In general, however, the two measures can differ, resulting in an inequivalent ordering imposed on the set of pure entangled quantum states $\ket\psi$ with respect to a bipartition involving at least a qu$d$it with $d \ge 4$. It is interesting to illustrate explicitly the discrepancy between the linear entropy and the stellar entanglement in the case $d=4$. In Fig.~\ref{figrandello} we report ${\cal E}_\sse$ versus $E_L$ for a sample of 20000 randomly generated states $\psi \in \C^{d_A} \otimes \C^{d_B}$ with $d_A=4$ and arbitrary $d_B \ge 4$ (upon applying the Schmidt decomposition, the effective dimension of each subsystem is reduced to $d=\min\{d_A,d_B\}$ which amounts to $4$ in this example). We find that, although physical qu$d$it states span a two-dimensional surface in the $(E_L,{\cal E}_\sse)$ plane, nonetheless for a given values of $E_L$ there exist sharp upper and lower bounds on ${\cal E}_\sse$. In fact, we will see below that these bounds are general and admit an exact analytical expression in any dimension. The classes of entangled states that saturate them in the case $d=4$ are specified in the caption of Fig.~\ref{figrandello}.

The pattern observed for the local dimension $d \le 4$ extends indeed to arbitrary values of the local dimension $d$. In particular, the hierarchical relationship ${\cal E}_\sse \le E_L$ always holds (with equality for local dimension $d=2,3$), and a rigorous proof is provided below. Moreover, the stellar entanglement $\sse$ME presents a structured, multi-branched {\it lower} bound as a function of the linear entropy of entanglement, with the number of branches growing with local dimension $d \geq 4$. Without aiming at a characterization of the complete lower boundary for states in Hilbert spaces of arbitrary dimension, we can nevertheless show that a simple rescaling of the linear entropy $E_L$ allows to derive an exact analytical lower bound on ${\cal E}_\sse$ that holds in general for arbitrary dimension. In the particular case of local dimension $d=4$ it corresponds to the bottommost branch of the lower boundary in Fig.~\ref{figrandello}. Indeed, one can prove that the following holds:

\begin{theorem}\label{Tlower}
The $\sse$ME is a lower bound to the linear entropy of entanglement and is an upper bound to a rescaling of it on all pure bipartite states in Hilbert spaces of arbitrary finite dimension:
\begin{equation}\label{eqlower}
\frac{2(d-1) \sin^2(\pi/d)}{d}E_L(\psi) \; \le {\cal E}_\sse(\psi) \; \le E_L(\psi) \; .
\end{equation}
\end{theorem}
The rightmost inequality is always tight and the states $\ket{\psi}$ that saturate it are those for which $|\sum_i\lambda_ip_i|$, with $\lambda_i$ being the eigenvalues of the stellar spectrum, \eq{stellarspec},
is invariant under permutations. Incidentally, we note that for local dimension $d=2,3$, all pure states have such permutational invariance. The proof of the rightmost inequality is provided in the Appendix~\ref{app:upperbound}.

The leftmost inequality is generally tight for $E_L \le d/[2(d-1)]$ and is saturated by states $\ket\psi$ with rank-2 marginal states $\varrho$: $p_{\sigma_1}=p$, $p_{\sigma_2}=1-p$, $p_{\sigma_i}=0$ ($i=3,\ldots,d$), for any local dimension $d$. For $d=2,3$, the lower and upper bounds in \eq{eqlower} coincide. The leftmost inequality is proven in Appendix~\ref{app:lowerbound}.

The attainable region for ${\cal E}_\sse$ at a fixed $E_L$ increases with increasing local dimension $d$, and the lower bound vanishes asymptotically as $d \rightarrow \infty$, showing that there can exist pure bipartite entangled states of two qu$d$its with local dimension $d \gg 1$, whose  linear entropy of entanglement lies in the range $0 < E_L \le \frac12$, and yet possess an infinitesimal degree of stellar entanglement $\sse$ME. The situation may change again and the trend may be reverted in the infinite-dimensional case. Indeed, in a previous study, some of us have shown that restricting to Gaussian pure bipartite states of two continuous variable modes, and to Gaussian LU operations, one can define a specific Gaussian counterpart to the stellar entanglement $\sse$ME that amounts to a simple monotonically increasing function of the linear entropy of entanglement \cite{squoCV}, and thus back in analogy to the case of low-dimensional qu$d$its ($d=2,3$) \cite{GiampaoloIlluminati}. In conclusion, the analysis reported in this Section shows that the stellar entanglement $\sse$ME, a prominent representative of the $\Lambda$ME mirror entanglement monotones, provides in general an independent characterization of bipartite entanglement in pure states of arbitrary finite dimension and is endowed with an intrinsic geometric origin depending entirely upon the global non-local effects induced only by the action of suitably identified classes of LU operations.

\section{Discussion and conclusion}\label{secDiscuss}

In the present work we have introduce a geometric framework to derive bipartite entanglement monotones, including faithful ones, that apply to all pure states in Hilbert spaces of arbitrary dimension. The measures of mirror entanglement and the faithful mirror stellar entanglement that we have introduced are defined in terms of the minimal (squared) distance from a pure state to the pure state obtained from it by the action of suitably optimized LUs acting only on one part of the bipartite quantum system under consideration.
We identified a hierarchy of these LU-based entanglement monotones, studied their properties, and determined conditions for their faithfulness. Among the faithful mirror entanglement measures, we focused on a special instance, the stellar mirror entanglement, characterized by additional symmetries in the spectrum of the associated LUs. We proved that the stellar mirror entanglement obeys upper and lower bounds as a function of the linear entropy of entanglement in any dimension, reducing to the latter for local dimension $d=2,3$.
Our results generalize an earlier study limited to pure states with reduced density matrices of lower local dimension $d=2,3$ \cite{GiampaoloIlluminati}.
Our work goes along a complementary direction compared to other studies that have investigated the nature of non-local effects when the distance from a state and its image under the action of LUs is maximized rather than minimized \cite{Bruss}. It is remarkable and somewhat surprising that looking at such a simple structure of LUs from opposite ends can provide so much insight both on the structure of pure state bipartite entanglement and, at the same time, on the patterns of non-local effects generated from the application of LU operations. This interplay, yielding a host of complementary results, might hide yet undiscovered features common to the two approaches. Both rely on the natural physical intuition of the
operational approach to the study of physical systems: looking at the response to a given action as basic diagnostic tool of physical properties. It might then be worth to try to compare the two different situations in terms of a classification/ordering of least-disturbing and/or maximally disturbing LUs.

An interesting and important subject for future research is the generalization of the entanglement structure generated by LUs to include mixed states. It has been recently shown \cite{Streltsov} that the problem of computing the convex roof of a prototype distance measure of entanglement such as the global geometric entanglement \cite{Wei} can in fact be recast in terms of determining the maximal fidelity to a separable state. It is tempting to speculate that this important result might perhaps be adapted and generalized to other classes of distance measures of entanglement, such as the mirror entanglement and the stellar entanglement introduced in the present work, or the mixed bipartite-multipartite geometric measures of entanglement defined as the hierarchy of distances from the sets of $K$-separable states \cite{Blasone}.

Extensions of the present investigation to the qualification and quantification of multipartite entanglement and the characterization of monogamy constraints on its distribution appear to be challenging, even when restricted to pure states. Here the problem appears of course to be that of understanding the nature of generalized "local operations" in the multipartite setting, and their ordering according to the associated local dimension. Viceversa, a more readily exploitable application of our results is concerned with the factorization of quantum ground states in cooperative qu$d$it systems, a phenomenon that is currently receiving significant attention from both quantum information and condensed matter communities \cite{faziorev,kurmannetal,ourfactoriz1,ourfactoriz2}. Indeed, the variety of quantum states belonging to one and the same LU-equivalence class may anyway have rather distinct features that become relevant especially in the context of many-body physics. For instance, if we consider a spin chain in a perfectly ferromagnetic state $\ket{\uparrow\uparrow\ldots\uparrow}$, flipping every other spin amounts to a LU operation that has no effect on the entanglement, yet results in a totally different ordered phase with a vanishing magnetization\footnote{In this context, we should also notice the fact that the states in the above example are fully factorized (fully separable), while the condensed matter terminology identifies them as strongly correlated, referring to the behavior of the two-point correlation functions; here we will adhere to the conventions of entanglement theory.}. The specific form of the ground state of a many-body Hamiltonian, then, is important as well as its entanglement content and distribution in the form of bipartite or multipartite quantum correlations \cite{faziorev}.
In this context, the formalism of LU-based geometric entanglement for states with reduced density matrices
of lower local dimension $d=2,3$ has already been applied to define energy witnesses
as efficient diagnostic tools of factorization, relating it to the vanishing of
faithful entanglement measures such as the stellar mirror entanglement. However, this
LU-based theory of ground-state factorization had to be restricted so far
to spin-1/2 systems \cite{giampiverruca} and to be limited to
investigate only total factorization into products of single-spin states \cite{ourfactoriz1,ourfactoriz2},
as the extension to higher spin systems and partial factorizations
(such as dimerization) required considering higher local
dimensions $d$. The results of the present work allow in principle to extend the LU-based
methods, originally developed for spin-1/2 models and total factorization,
to the study of total as well as partial factorization points in models of higher-dimensional spins
and in spin-1/2 models with frustration.
In the latter case of frustrated spin-1/2 systems, the LU-based entanglement
formalism has been recently exploited to relate the existence of fully separable
ground states (totally factorized in the tensor product of single-spin states)
to the absence of frustration \cite{ourfrust}.
Equipped with the general proof of equivalence between pure-state factorization and invariance under LU
spin operations with stellar (or, generally non-degenerate) spectrum, one may now investigate the
possible existence of dimerized quantum phases (i.e.~ground states composed of factorized singlets: tensor products of $d=4$ units, each of which is internally entangled) by looking for candidate ground
states invariant under stellar LUs in $d=4$. For spin-1/2 chains in the thermodynamics limit or
in $2D$ lattices one may expect that a hierarchy of compatible types of ground state
correlations may take place, ranging from full factorization into products of single-spin states,
to dimerization, up to genuine multipartite-entangled phases, as the degree of
frustration increases as a function of a given tunable external magnetic field.
The exploration of these intriguing scenarios, as enabled by the general analysis developed
in the present work, will be the subject of further future studies.

\acknowledgments{F.I. acknowledges useful discussions with Dagmar Bru{\ss}. A.M. acknowledges useful discussions with J. I. de Vicente. We are particularly grateful to Marco Piani, who pointed out to us
that an early proof of the monotonicity of the mirror entanglement under LOCC was incomplete, and
prompted us to derive the complete and correct proof that is reported in the present work.
We finally thank the European Commission of the European Union for financial support under the FP7 STREP Project HIP (Hybrid Information Processing), Grant Agreement n. 221889.}

\appendix
\section{Proof of ${\cal E}_\sse(\psi) \le E_L(\psi)$}\label{app:upperbound}
In order to prove the theorem, we will use Dirac bracket notation throughout. It is convenient to establish some definitions and facts. Let us denote the maximally mixed vector by $\ket\openone=(1/d,1/d,\ldots,1/d)^\dg$. With this, $\langle\openone|\openone\rangle=1/d$ and we denote the projector onto the $\ket\openone$ subspace by $P^\parallel=d\ket\openone\bra\openone$, the complement being $P^\perp=\openone-P^\parallel$. The eigenvalues $\lambda_i$ can be arranged in the vector $\ket\lambda$, with complex conjugate $\ket{\lambda^*}$. We have $\langle\lambda|\lambda\rangle=\langle \lambda^*|\lambda^*\rangle=d$ and for the stellar monotone, also $\langle\lambda|\openone\rangle=0$.

Define $M_\sigma=\frac{1}{2}\sigma^\top (\ket\lambda\bra\lambda+\ket{\lambda^*}\bra{\lambda^*})\sigma$.  Denoting the identity permutation by $e$, we can write $M\equiv M_e=\Re\,\ket\lambda\bra\lambda$, and $M_\sigma=\sigma^\top M\sigma$. It is clear that $M\geq0$ and thus $M_\sigma\ge0$ for all $\sigma$. Let
\begin{equation}
	g_\sigma(p)=1-\big|\sum_i\lambda_i p_{\sigma_i}\big|^2=1-\bra pM_\sigma\ket p.
\end{equation}
 For any $0< s\leq 1$, let ${\cal G}_{e,s} =\{p|g_e(p) \geq s\}$. Proceed now to define ${\cal C}_s=\{p~|~{\cal E}_\sse(p)\geq s\}$. Then ${\cal C}_s=\cap_{\sigma\in S_d}{\cal G}_{s,\sigma}$.

Before proceeding any further, we will find the following relation useful,
\begin{equation}
	\sum_{\sigma\in\textsf{S}_d}M_\sigma=d!\frac{d}{d-1}P^\perp,
\end{equation}
which can be easily derived by use of Schur's lemma.

%

The proof consists on proving a lower bound of $E_L$ over the set ${\cal C}_{s}$. Equivalently, we may search for the minimum of the objective function $f(p)=1-\langle p|p\rangle$,
\begin{align}\label{eq:program}
	\textrm{minimize~}& f(p)\\
	\nonumber
	\textrm{subject to}~&p\in{\cal C}_s.
\end{align}
Let us denote the solution to this problem by $p^*$ and the achieved value $f(p^*)$ by $f^*_s=f(p^*)$.\\

\noindent {\bf Lemma}: The solution to the optimization in \eqref{eq:program} satisfies
\begin{equation}
	f^*_s\leq s\frac{d-1}{d}.
\end{equation}
\noindent
{\em Proof}. Consider the vector $\ket q=(1-\sqrt{1-s})\ket\openone+\sqrt{1-s}\ket{e_1}$, where $\ket{e_1}=(1,0,0,\ldots)^\dg$. This vector has $g_\sigma(q)=s$, for all $\sigma$, thus $q\in{\cal C}_s$, and $f(q)=s\frac{d-1}{d}$. Therefore $f^*_s\leq f(q)=s\frac{d-1}{d}$\hfill$\Box$\\*

\noindent {\bf Lemma}: The solution to the optimization in \eqref{eq:program} satisfies
\begin{equation}
	f^*_s\geq s\frac{d-1}{d}.
\end{equation}
\noindent
{\em Proof}. The program \eqref{eq:program} is spelled out as
\begin{align}\nonumber
	\textrm{minimize~~~}& f(p)\\
\label{eq:primal}
	\textrm{subject to~~~}&\left\{
	\begin{aligned}
		s-g_\sigma(p)&\leq0\\
					   -p_i&\leq0\\
					   \sum_ip_i-1&=0
	\end{aligned}\right.,
\end{align}
and the Lagrangian is
\begin{multline}
	L(p,\mu,\alpha,\nu)=\\
	f(p)+\sum_\sigma \mu_\sigma(s-g_\sigma(p))-\langle \alpha|p\rangle-\nu(1-d\langle \openone|p\rangle).
\end{multline}
Consider the dual program~\cite{boyd_convex,convexanalysis} to that of \eqref{eq:primal}. The dual Lagragian $L_D$ is
\begin{equation}\label{eq:dual}
	L_D(\mu,\alpha,\nu)=\inf_{p}L(p,\mu,\alpha,\nu) \, .
\end{equation}
In order to have the dual Lagrangian $L_D(\mu_\sigma,\alpha,\nu)>-\infty$ it is convenient to write the objective function as
\begin{equation}
	f(p)=\bra p(d^2\ket \openone\bra\openone-\openone)\ket p.
\end{equation}
$L(p,\mu,\alpha,\nu)$ can then be written as
\begin{align}
	L(p,\mu,\alpha,\nu)=&\,\bra p\left(d^2\ket \openone\bra\openone-\openone+\sum_\sigma\mu_\sigma M_\sigma\right)\ket p\\
	\nonumber
	&-\left(\bra{\alpha}-\nu d\bra\openone\right)\ket p-(1-s)\sum_\sigma\mu_\sigma-\nu.
\end{align}
Hence, $L$ is bounded from below whenever $H\equiv\nabla^2_p L=\sum_\sigma\mu_\sigma M_\sigma-\openone>0$. In such case, the infimum in~\eqref{eq:dual} can be readily computed by solving $\nabla_p L=0$,
\begin{equation}
	\ket p=\frac{1}{2}H^{-1}\left[\ket\alpha-\nu d\ket\openone\right],
\end{equation}
yielding
\begin{align}\nonumber
	L_D(\mu,\alpha,\nu)=&-\frac{1}{4}\left(\bra{\alpha}-\nu d\bra\openone\right)H^{-1}\left(\ket\alpha-\nu d\ket\openone\right)\\
	&-(1-s)\sum_\sigma\mu_\sigma-\nu \, .
\end{align}
At this point, any value of $L_D$ with $\{\mu_\sigma\}$ satisfying $H>0$ yields a lower bound to $f^*(s)$. Moreover, observe that $\sum_\sigma\mu_\sigma=\tr[H]/d$,
\begin{align}\nonumber
	L_D(\mu,\alpha,\nu)=&-\frac{1}{4}\left(\bra{\alpha}-\nu d\bra\openone\right)H^{-1}\left(\ket\alpha-\nu d\ket\openone\right)\\
	&-(1-s)\tr[H]/d-\nu.
\end{align}

We can choose to evaluate $L_D(\mu,\alpha,\nu)$ at $\mu_\sigma=\mu$ so that $H$ becomes
\begin{equation}
	H=\left(\mu\frac{d!d}{d-1}-1\right) P^\perp+(d-1)P^\parallel \, .
\end{equation}
Defining $x\equiv \mu\frac{d!d}{d-1}-1$, we have
\begin{subequations}
\begin{eqnarray}
	H&=&x P^\perp+(d-1)P^\parallel\\
	H^{-1}&=&x^{-1}P^\perp+(d-1)^{-1}P^\parallel \, ,
\end{eqnarray}
\end{subequations}
while the $H> 0$ condition reduces to $x>0$ or $\mu>\frac{d-1}{d!d}$. Moreover, $\sum_\sigma \mu_\sigma=\tr[H]/d=(1+x)\frac{d-1}{d}$. Thus, we obtain
\begin{align}
\nonumber
	L_D(x,\alpha,\nu)=&-\frac{1}{4x}\|\ket{\alpha^\perp}\|^2-(1-s)\frac{d-1}d x\\
\nonumber
	&-\frac{1}{4(d-1)}\|\ket{\alpha^\parallel}-\nu d\ket\openone \|^2\\
	&-(1-s)\frac{d-1}{d}-\nu.
\end{align}
Maximizing over $x$ we obtain $x^*=\frac{\|\ket{\alpha^\perp}\|}{2\sqrt{1-s}}\sqrt{\frac{d}{d-1}}$ and
\begin{eqnarray}\nonumber
	L_D(x^*,\alpha,\nu)&=&-\|\ket{\alpha^\perp}\|\sqrt{(1-s)\frac{d-1}{d}}\\
\nonumber
	&&-\frac{1}{4(d-1)}\|\ket{\alpha^\parallel}-\nu d\ket\openone \|^2\\
	&&-(1-s)\frac{d-1}{d}-\nu.
\end{eqnarray}
The term proportional to $\|\ket{\alpha^\perp}\|$ can only be negative, thus we set it to zero by choosing $\ket\alpha=\ket{\alpha^\parallel}=\alpha \sqrt d\ket\openone$ ($\alpha\geq0$), leaving
\begin{eqnarray}
	L_D(x^*,\alpha,\nu)=-\frac{(\alpha-\nu\sqrt d)^2}{4(d-1)}-\nu-(1-s)\frac{d-1}{d}.~~~
\end{eqnarray}
Maximizing over $\nu$ we get $\nu^*=2\frac{\alpha}{\sqrt d}-4\frac{d-1}{d}$ and
\begin{equation}
	L_D(x^*,\alpha,\nu^*)=s\frac{	d-1}{d}-\frac{\alpha}{\sqrt d}.
\end{equation}
Since $\alpha\geq0$ the maximum is achieved at $\alpha^*=\alpha=0$ yielding
\begin{equation}
	L_D(x^*,\alpha^*,\nu^*)=s\frac{d-1}{d} \, ,
\end{equation}
which shows that
\begin{equation}
	f^*_s\geq s\frac{d-1}{d}.
\end{equation}
\hfill$\Box$

Combining the last two lemmas, we have
\begin{equation}
	f^*_s=s\frac{d-1}{d}.
\end{equation}
From this we have the following relation,
\begin{equation}
	{\cal E}_\sse(p)\geq s\Rightarrow f(p)\geq s\frac{d-1}{d},
\end{equation}
or equivalently
\begin{equation}
	{\cal E}_\sse(p)\geq s\Rightarrow E_L(p)\geq s.
\end{equation}
In particular, for points with ${\cal E}_\sse(p)=s$ we have
\begin{equation}
	{\cal E}_\sse(p)=s\leq E_L(p) \, .
\end{equation}
This completes the proof of the rightmost inequality in Theorem \ref{Tlower}.\hfill$\Box$

\section{Proof of $2\frac{d-1}d \sin^2(\pi/d)E_L(\psi) \le {\cal E}_\sse(\psi)$}\label{app:lowerbound}
For this proof we will use the same notation as in Appendix~\ref{app:upperbound}. Additionally, let $\Lambda=\ket\lambda\bra\lambda$. The approach of the proof is the following. Since all entanglement monotones can be expressed as function of the spectrum of the reduced density matrix, we will talk about {\em probability vectors} instead of quantum states. We will show that any probability vector $\ket p$ can be obtained by a series of transformations starting from the pure vector $\ket q=(1,0,0\ldots)^\dg$ such that the increments in both entanglement monotones under the action of these transformations always verify
\begin{equation}\label{eq:incineq}
	\Delta {\cal E}_\sse\geq 2\frac{d-1}d \sin^2(\pi/d)\Delta E_L.
\end{equation}
This, combined with the fact that both monotones vanish on $(1,0,0,\ldots)$ completes the proof. We now proceed to prove Eq.~\eqref{eq:incineq}. An essential ingredient in the proof consists in obtaining a sequence of transformations that will bring $q$ to $p$ while still having a manageable form of $\Delta {\cal E}_\sse$. To this end, we realize that the most disturbing element in the expression of $\Delta{\cal E}_\sse$ are the optimizations over the set of permutation matrices for the initial and final states. We can avoid this complication by showing that a sequence of transformations can be devised such that for each one of them, the initial and final states have at least one optimal permutation matrix in common. This will allow us to get rid of the independent optimization for the initial and final values ${\cal E}_\sse(\psi)$.

Any quantum state with reduced density matrix $\varrho$ whose eigenvalues are $p_i$ is majorized by a pure state with $q=(1,0\ldots)$, which means that
\begin{equation}\label{eq:ttransf}
	\ket p=T_{d-1}\cdots T_2T_1\ket q \, ,
\end{equation}\\
where the $T_k$'s are T-transforms [{\em i.e.}, matrices of the form $T(t)=(1-t) \openone+t W$ where $W$ is a transposition of two particular elements and $0\leq t\leq 1$], and $t_k$ are their respective arguments. We will say that $T(t)=(1-t)\openone+t W$ is a T-transform {\em of the kind} $W$. T-transforms with $t\geq1/2$ can be reduced to T-transforms with $ t\leq 1/2$ by prepending the $W$ permutation: $T(t)W=T(1-t)$. Thus, Eq.~\eqref{eq:ttransf} can be cast as
\begin{equation}\label{eq:ttransf2}
	\ket p=T_{d-1}W_{d-1}\cdots T_2 W_2T_1W_1\ket q \, ,
\end{equation}\\
where $W_1,W_2,\ldots$ are appropriate permutation matrices (could be the identity) and now all T-transforms have arguments $t_k\leq1/2$.

 Moreover, any T-transform with $0\leq t\leq1/2$ can be split into an arbitrary number of intermediate transforms of the same kind $T(t'),T(t''),\ldots$ with $0\leq t',t'',\ldots\leq1/2$. This can be seen by re-parameterizing the set of T-transforms of any given kind [$0\leq t\leq1/2$] by
\begin{equation}
	T(s)=\frac{1+e^{-s}}{2} \openone+\frac{1-e^{-s}}2W \, ,
\end{equation}
where $T(0)=\openone$ and $0\leq s<\infty$. Observing that $T(s_1)T(s_2)=T(s_1+s_2)$ shows that any splitting of $s=\sum_i s_i$ with positive summands can be accomplished. This allows to decompose each $T_k$ into a product of $T_k^{n_k}\cdots T_k^2T_k^1$ such that the states $(T_k^{i-1}\cdots T_k^1)\,T_{k-1}\cdots T_2 T_1\ket q$ and $(T_k^iT_k^{i-1}\cdots T_k^1)\,T_{k-1}\cdots T_2 T_1\ket q$ share at least one optimal permutation matrix, for all $i,k$. Thus, finally the target state $\ket p$ can be written as
\begin{equation}
	\ket p= \prod_{k=1}^{d-1}\left[\left(\prod_{i=1}^{n_k} T_k^i \right)W_k\right]\ket q \, .
\end{equation}

We are now ready to show that Eq.~\eqref{eq:incineq} holds for any transformation $\ket p\rightarrow T_k^l \ket p$. Let $\sigma$ be a common optimal permutation matrix for both states, and $\Lambda_\sigma=\sigma^\top\Lambda\sigma$, and $T_k^l=(1-t) \openone+tW$, with $t\leq1/2$. The increment in ${\cal E}_\sse$ can be cast as
\begin{align}\nonumber
	\Delta &{\cal E}_\sse(p)\\
	\nonumber
	=&\,\bra p \Lambda_\sigma\ket p-\bra p T(t)^\top\Lambda_\sigma T(t)\ket p\\
	\nonumber
	=&\,\bra p\left((1-(1-t)^2)\Lambda_\sigma-t(1-t)\left[\Lambda_\sigma W+W\Lambda_\sigma\right]\right)\ket p\\
	\nonumber
	&\,-t^2\bra pW\Lambda_\sigma W\ket p\\
\label{eq:bound1}
	\geq&\,2 t(1-t)\bra p \left[\Lambda_\sigma-\frac{1}{2}(W\Lambda_\sigma+\Lambda_\sigma W)\right]\ket p.
\end{align}
In the last inequality we have exploited the assumption that $\sigma$ is the optimal permutation, hence $\bra p\Lambda_\sigma\ket p\geq \bra pW\Lambda_\sigma W\ket p$.

Evaluating expression \eqref{eq:bound1} we find
\begin{equation}
	\bra p \left[\Lambda_\sigma-\frac{1}{2}(W\Lambda_\sigma+\Lambda_\sigma W)\right]\ket p=2\Delta p^2\sin^2\frac{\Delta\theta}{2},
\end{equation}
where $\Delta p^2=(p_i-p_j)^2$ where $i,j$ are the indices swapped by $W$, or the subspace where $T_k^l$ acts non-trivially, and $\Delta \theta=\theta_{\sigma^{-1}_i}-\theta_{\sigma^{-1}_j}=\frac{2\pi}{d}(\sigma^{-1}_i-\sigma^{-1}_j)$. Indeed, since
\begin{equation}
	\sin^2\frac{\Delta\theta}{2}\geq\sin^2\frac{\pi}{d} \; ,
\end{equation}
we have
\begin{equation}
	\Delta {\cal E}_\sse(p)\geq4 t(1-t)\Delta p^2\sin^2\frac{\pi}{d} \; .
\end{equation}

It is then finally straightforward to verify that $4t(1-t) \Delta p^2=2\frac{d-1}{d}\Delta E_L$. \hfill$\Box$

\end{document}